\documentclass[aps,prl,twocolumn,groupedaddress,showpacs]{revtex4-1}
\usepackage{dcolumn} 
\usepackage{bm} 
\usepackage{amsthm}
\usepackage{latexsym}
\usepackage{float}
\usepackage{amsfonts}
\usepackage{amsmath}
\usepackage{amssymb}
\usepackage{color,graphicx}
\usepackage{booktabs}
\usepackage{epstopdf}

\begin{document}
\title{Detecting electron-phonon couplings during photo-induced phase transition}
\author{Takeshi~Suzuki$^{1,*}$}
\author{Yasushi~Shinohara$^{2,3}$}
\author{Yangfan Lu$^{4}$}
\author{Mari Watanabe$^{1}$}
\author{Jiadi Xu$^{1}$}
\author{Kenichi L. Ishikawa$^{2,3,5}$}
\author{Hide Takagi$^{4,6}$}
\author{Minoru Nohara$^{7}$}
\author{Naoyuki Katayama$^{8}$}
\author{Hiroshi Sawa$^{8}$}
\author{Masami Fujisawa$^{1}$}
\author{Teruto Kanai$^{1}$}
\author{Jiro Itatani$^{1}$}
\author{Takashi Mizokawa$^{9}$}
\author{Shik Shin$^{10,11,*}$}
\author{Kozo Okazaki$^{1,11,12,*}$}
\affiliation{$^1$Institute for Solid State Physics, The University of Tokyo, Kashiwa, Chiba 277-8581, Japan\\
$^2$ Photon Science Center, Graduate School of Engineering, The University of Tokyo, 7-3-1 Hongo, Bunkyo-ku, Tokyo 113-8656, Japan\\
$^3$ Department of Nuclear Engineering and Management, Graduate School of Engineering, The University of Tokyo, 7-3-1, Hongo, Bunkyo-ku, Tokyo 113-8656, Japan\\
$^4$ Department of Physics, University of Tokyo, Hongo, Tokyo, 113-0033, Japan\\
$^5$ Research Institute for Photon Science and Laser Technology, The University of Tokyo, 7-3-1 Hongo, Bunkyo-ku, Tokyo 113-0033, Japan\\
$^6$Max Planck Institute for Solid State Research, Heisenbergstrasse 1, 70569 Stuttgart, Germany \\
$^7$ Research Institute for Interdisciplinary Science, Okayama University, Okayama, 700-8530, Japan \\
$^8$ Department of Applied Physics, Nagoya University, Nagoya 464-8603, Japan \\
$^9$ School of Advanced Science and Engineering, Waseda University, Shinjuku, Tokyo 169-8555, Japan \\
$^{10}$ Office of University Professor, The University of Tokyo, Kashiwa, Chiba 277-8581, Japan \\
$^{11}$ Material Innovation Research Center, The University of Tokyo, Kashiwa, Chiba 277-8561, Japan \\
$^{12}$ Trans-scale Quantum Science Institute, The University of Tokyo, Bunkyo-ku, Tokyo 113-0033, Japan
}

\date{\today}

\begin{abstract}
Photo-induced phase transitions have been intensively studied owing to the ability to control a material of interest in the ultrafast manner, which can induce exotic phases unable to be attained at equilibrium.
However, the key mechanisms are still under debate, and it has currently been a central issue how the couplings between the electron, lattice, and spin degrees of freedom are evolving during photo-induced phase transitions.
Here, we develop a new analysis method, frequency-domain angle-resolved photoemission spectroscopy, to gain precise insight into electron-phonon couplings during photo-induced insulator-to-metal transitions for Ta$_2$NiSe$_5$.
We demonstrate that multiple coherent phonons generated by displacive excitations show band-selective coupling to the electrons.
Furthermore, we find that the lattice modulation corresponding to the 2 THz phonon mode, where Ta lattice is sheared along the a-axis, is the most relevant for the photo-induced semimetallic state.
\end{abstract}

\maketitle



\begin{figure*}[!t]
\includegraphics{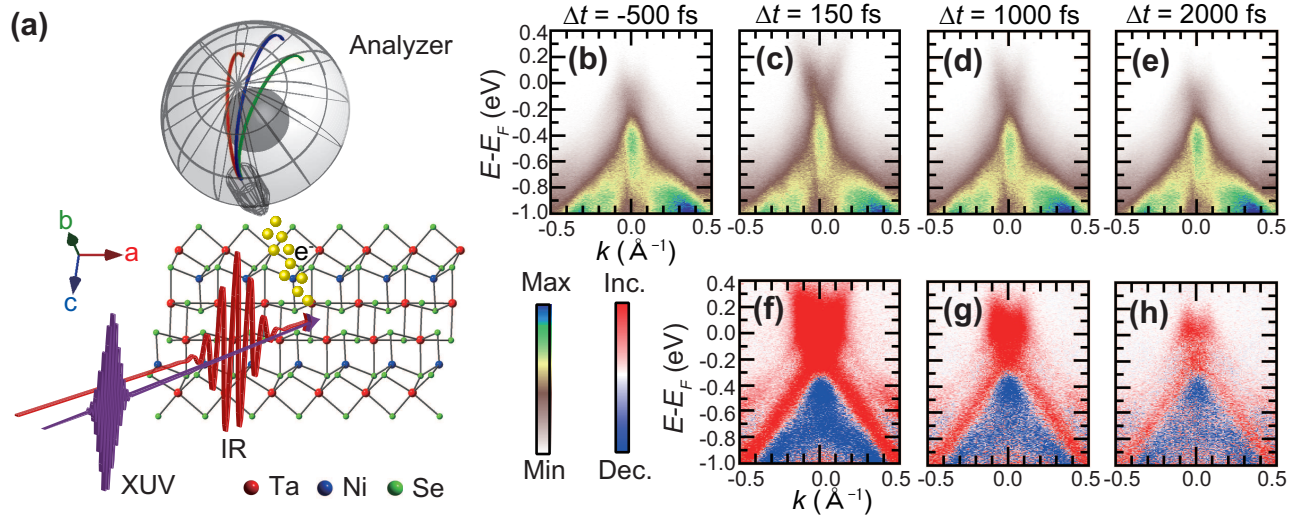}
\caption{
(a) Schematic illustration of time- and angle-resolved photoemission spectroscopy (TARPES), as applied to Ta$_2$NiSe$_5$.
The pump pulse is infrared light whereas the probe pulse is extreme ultraviolet light produced by high-harmonic generation.
Photoelectrons are detected by a hemisphere analyzer.
(b)-(e) TARPES spectra of Ta$_2$NiSe$_5$.
The delay time between the pump and probe is indicated in each panel.
(f)-(h) Difference images of TARPES.
Red and blue points represent increasing and decreasing photoemission intensity, respectively.
}
\label{fig1}
\end{figure*}

Strongly-correlated electron systems display very rich phases owing to intertwined couplings between multiple degrees of freedom including the charge, orbital, spin, and lattice \cite{Tokura_2017}.
Moreover, external fields, such as electronic and magnetic fields or physical pressure, can induce phase transitions in these systems by breaking their subtle balances between multiple competing phases \cite{Sow_2017, Matsuura_2017}.
In this respect, photo-excitation is a very promising way to control the physical properties because it can instantaneously change physical properties of a targeting material in various manners by exploiting many degrees of freedom such as polarization or wavelength \cite{Basov_2017}.
For studying photo-excited nonequilibrium states, time- and angle-resolved photoemission spectroscopy (TARPES) has a strong advantage because it can track nonequilibrium electronic band structures after photoexcitations \cite{Rohwer_2011, Hellmann_2012, Schmitt_2008, Suzuki_2019}.

For photo-induced phase transitions, although many strongly-correlated electron systems have been intensively studied \cite{Miyano_1997, Collet_2003, Fausti_2010, Frigge_2017, Zong_2019}, the precise mechanisms are still under debate \cite{Gedik_2007, Morrison_2014, Ichikawa_2011}.
Recently, we revealed the photo-induced insulator-to-metal transitions (IMTs) in Ta$_2$NiSe$_5$ \cite{Okazaki_2018}, where we also showed strong evidence in dynamical behaviors as an excitonic insulator.
Moreover, other interesting photo-excited phenomena in Ta$_2$NiSe$_5$ have been reported previously, which include photo-induced enhancement of the excitonic insulator \cite{Mor_2017, Tanabe_2018} or emergence of collective modes \cite{Murakami_2017, Werdehausen_2018}.
In most reports, the key roles for such phenomena are played by significant electron-phonon couplings.

In this Letter, we report an analysis method extended from TARPES, namely, frequency-domain angle-resolved photoemission spectroscopy (FDARPES), to reveal how electron-phonon couplings play roles in the photo-excited nonequilibrium states for Ta$_2$NiSe$_5$.
We observe that the lattice modulation corresponding to the phonon mode, where Ta lattice is sheared along the a-axis, is the most relevant for the photo-induced semimetallic state.
Furthermore, this method can be generally applicable to detect the temporal modulations of photoemission intensity induced by any coupling beyond coherent phonons.

TARPES, as illustrated in Fig. 1(a), allows us to directly observe the temporal evolution of the electronic band structure.
We used an extremely stable commercial Ti:sapphire regenerative amplifier system (Spectra-Physics, Solstice Ace) with a center wavelength of 800 nm and pulse width of $\sim$35 fs for the pump pulse.
Second harmonic pulses generated in a 0.2-mm-thick crystal of $\beta$-BaB2O4 were focused into a static gas cell filled with Ar to generate higher harmonics.
By using a set of SiC/Mg multilayer mirrors, we selected the seventh harmonic of the second harmonic ($h \nu$  = 21.7 eV) for the probe pulse.
The temporal resolution was determined to be $\sim$70 fs from the TARPES intensity far above the Fermi level, which corresponded to the cross correlation between the pump and probe pulses.
The hemispherical electron analyzer (Omicron-Scienta R4000) is used to detect photoelectrons.
All the measurements in this work were performed at the temperature of 100 K.

The sample is a high-quality single crystals of Ta$_{2}$Ni(Se$_{0.97}$S$_{0.03}$)$_{5}$ grown by chemical vapour transport method using I$_2$ as transport agent, as was reported in the refs \cite{Lu_2017, Nakano_2018}.
Whereas a relatively large cleaved surface is necessary for TARPES measurements compared with static ARPES, because Ta$_2$NiSe$_5$ has a one-dimensional crystal structure, a large cleaved surface of the pristine Ta$_2$NiSe$_5$ that was sufficient for TARPES measurements was difficult to obtain.
However, sufficiently-large cleaved surface of 3 \% S-substituted Ta$_2$NiSe$_5$ could be obtained.
This is why we used 3 \% S-substituted Ta$_2$NiSe$_5$ rather than pristine Ta$_2$NiSe$_5$ in this study.
Clean surfaces were obtained by cleaving in situ.

Figures 1(b)-1(e) show the TARPES snapshots of Ta$_2$NiSe$_5$ at various delays shown as a function of momentum and energy.
The pump fluence is set to be 2.27 mJ/cm$^{2}$.
The delay between the pump and probe is indicated in each panel.
To enhance the temporal variations, the difference images between the before and after photoexcitation are shown in Figs. 1(f)-1(h), where red and blue represent an increase and decrease in photoemission intensity, respectively.
After strong photoexcitation, new semimetallic electron- and hole-dispersions appear, as has previously been reported \cite{Okazaki_2018}.
This is the direct signature of photo-induced IMT.
To highlight the change of the electronic band structure, we show the peak positions of the TARPES spectra before and after photoexcitation in Figs. 2(a) and 2(b).
One can notice that the hole band is shifted upward and crosses the Fermi level, $E_\text{F}$, while the electron band appears and crosses $E_\text{F}$ at the same Fermi wavevector, $k_\text{F}$, as the hole band.

\begin{figure}[!b]
\includegraphics{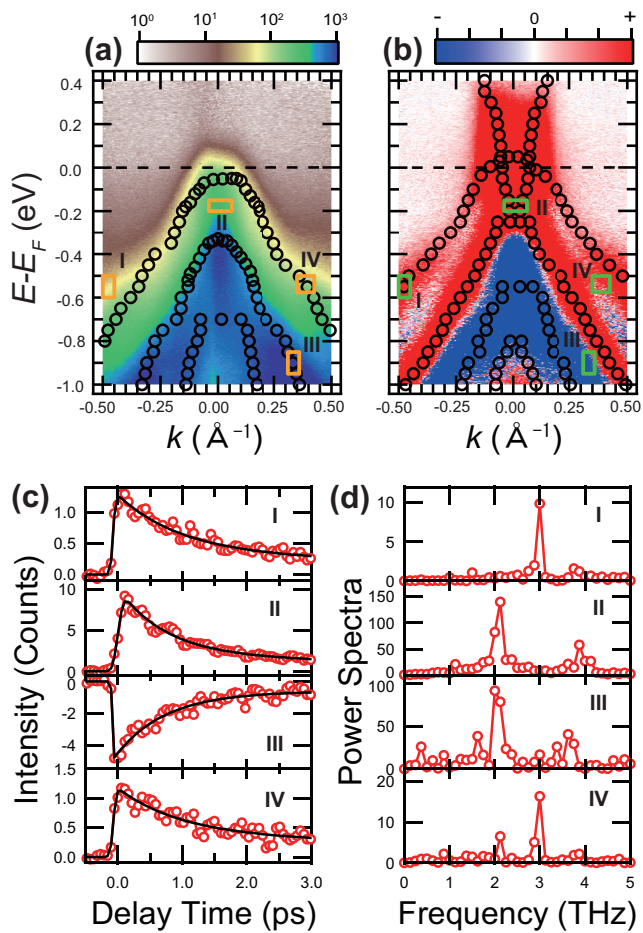}
\caption{
TARPES spectra (a)before and (b)after pump excitation.
The peak positions in the TARPES spectra are indicated as circles.
(c) Time-dependent TARPES intensities at different energy and momentum regions.
I-IV corresponds to the regions indicated in (a) and (b).
Data are shown as red circles whereas the fitting results by double-exponential decay functions convoluted with a Gaussian function are shown as black solid lines.
(d) Amplitude of Fourier transforms of the oscillation components in (c)I-(c)IV obtained by subtracting the fitting curve from the data.
}
\label{fig2}
\end{figure}

\begin{figure}[!t]
\includegraphics{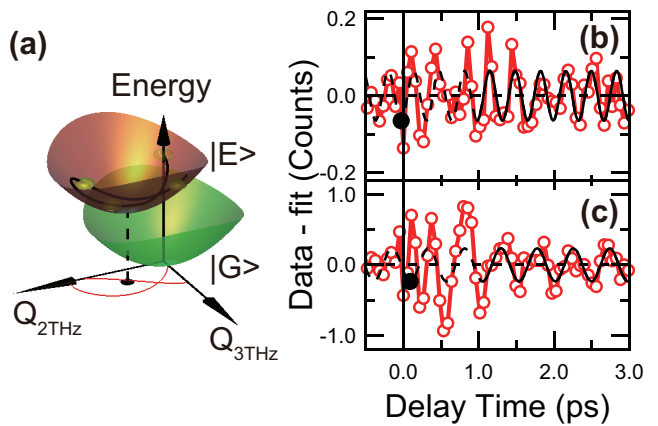}
\caption{
(a) Schematic illustration of the free energy curves for the ground and photo-excited states as a function of lattice coordinates corresponding to the directions of the 2-THz and 3-THz phonon modes.
(b), (c) Oscillation components for I and II corresponding to the data subtracted by the fits.
The fitting curve using a single cosine function are shown as dotted and solid black curves.
We use data for the fits shown by solid black curves. 
}
\label{fig3}
\end{figure}

To more specifically reveal the photo-induced profile in Ta$_2$NiSe$_5$, we investigated the TARPES images in terms of electron-phonon couplings.
Figure 2(c) shows the time-dependent intensities for representative regions in the energy and momentum space indicated as I-IV in Figs. 2(a) and 2(b).
As a background, carrier dynamics corresponding to overall rise-and-decay or decay-and-rise behaviors were observed.
Additionally, oscillatory behaviors were clearly seen superimposed onto background carrier dynamics, which indicated strong electron-phonon couplings as a result of excitations of coherent phonons.
To extract the oscillatory components, we first fit the carrier dynamics to a double-exponential function convoluted with a Gaussian function, shown as the black-solid lines in Fig. 2(c), and then subtracted the fitting curves from the data.
Fourier transformations were performed for the subtracted data and the intensities for each frequency component are shown in Fig. 2(d).
One can clearly see that distinctively different frequency-dependent peak structures appeared depending on the regions in the energy and momentum space denoted as I-IV in Figs. 2(a) and 2(b).

\begin{figure*}[!t]
\includegraphics{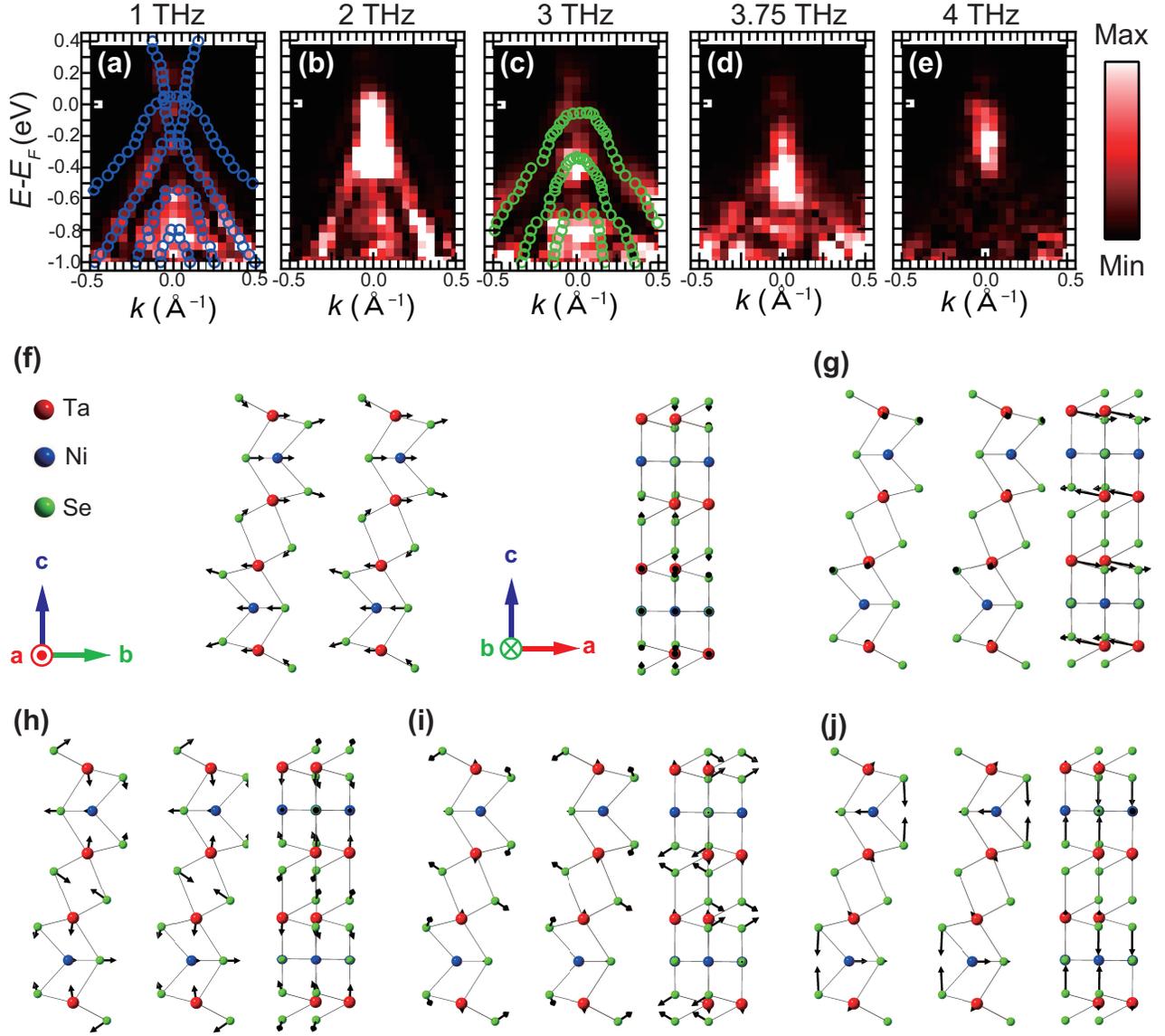}
\caption{
(a)-(e) Frequency-domain angle-resolved photoemission spectroscopy (FDARPES) spectra shown as frequency-dependent intensities of the oscillation components as a function of energy and momentum.
The peak positions in the TARPES spectra after and before photoexcitation are plotted as blue and green circles in Fig. 4(a) and Fig. 4(c), respectively.
(f)-(j) Calculated phonon modes corresponding to (a)-(e).
}
\label{fig4}
\end{figure*}

Considering that the frequencies for the peak positions observed in this work matched the $A_g$ phonon modes reported in previous work \cite{Kim_2016, Mor_2018} and that we dominantly excited the electron system under our experimental condition, the observed coherent phonons were likely to arise from displacive excitation of coherent phonons (DECPs) \cite{Zeiger_1992}.
According to the DECP theory, photoexcitation suddenly changes the minimum energy position in the lattice coordinates of the potential energy surface (PES), and lattice coordinates oscillate around the new energy minimum position with its own frequency determined by the curvature of the PES.
Figure 3(a) schematically shows this situation.
The ground and excited states are denoted as $\left| G \right>$ and $\left| E \right>$, respectively, while $Q_{\text{2THz}}$ and $Q_{\text{3THz}}$ are the lattice coordinates corresponding to the 2-THz and 3-THz phonon modes.

To investigate the phases of coherent phonons, we further fitted the oscillatory components by single-cosine function.
Figures 3(b) and 3(c) shows the results for regions I and II, in which the 3- and 2-THz components exhibited the strongest peaks, respectively.
Phases and frequencies were obtained as -0.19$\pm$0.11 $\pi$ and 2.97$\pm$0.02 THz for region I, and 0.37$\pm$0.12 $\pi$ and 2.07$\pm$0.02 THz for region II, respectively.
If the change of PES accompanies no delay from the laser excitation, DECP has a cosine-like behavior, that is, the expected phase is 0 $\pi$.
Thus, the modulation of photoemission intensity in region I along the 3-THz phonon mode was triggered immediately after photoexcitation.
On the other hand, the relatively positive phase shift in region II compared with region I indicated the modulation of the photoemission intensity along the 2-THz phonon mode occurred with a delay of 120 fs.

We will now discuss the electron-phonon couplings in more detail.
Since we observed that the amplitude of each oscillation significantly changed depending on the regions in the energy and momentum space, we further mapped out the frequency-dependent intensity of the Fourier component in the energy and momentum space, which we call FDARPES.
Figures 4(a)-4(e) show the FDARPES spectra corresponding to the frequencies of 1, 2, 3, 3.75, and 4 THz, respectively.
The corresponding phonon modes were calculated by ab initio calculation and their modes are shown in Figs. 4(f)-4(j).
The details of calculations are found in Supplemental Material \cite{Suppl}.
We assigned all the phonon modes as $A_g$ modes in the monoclinic phase.
In the previous results of Raman measurements at different polarization settings \cite{Werdehausen_2018}, eleven Raman-active phonon modes are observed in the Y(ZZ)Y setting while the three of them show stronger peak intensities in the Y(ZX)Y than in the Y(ZZ)Y.
This is considered to be due to the fact that all of the observed modes have $A_g$ character in the monoclinic phase and three of them turn to be $B_{2g}$ character in the orthorhombic phase.
Precise procedures of phonon assignments are found in Supplemental Material \cite{Suppl}.

Noticeably, the FDARPES spectra exhibit distinctively different behaviors depending on the frequency, which demonstrates that each phonon mode is selectively coupled to the specific electronic bands.
Particularly, the 2-THz phonon mode has the strongest signal around $E_\text{F}$, where it consists of a mixture of Ta $5d$ and Se $3p$ orbitals \cite{Lee_2019}, and this signature is responsible for the collapse of the excitonic insulator.
Because the IMT of Ta$_2$NiSe$_5$ occurs by melting the excitonic insulating flat band, this strongest signal suggests that the photo-induced metallic states drive the lattice distortions corresponding to the 2-THz phonon motion.

In order to see spectral features of FDARPES in more detail, we compare the FDARPES spectra with the band dispersions before and after photoexcitation.
Full results are found in Supplemental Material \cite{Suppl}.
We find that FDARPES spectrum at 1 THz matches the band dispersions after photoexcitation better than that before photoexcitation shown in Fig. 4(a) while the FDARPE spectrum at 3 THz is closer to the band dispersions before photoexcitation shown in Fig. 4(c).
Recent theoretical investigation reported that the intensity of FDARPES spectra reflects the strength of electron-phonon coupling matrix elements \cite{Giovannini_2020}.
Our results suggest the strong electron-phonon couplings for 1- and 3-THz phonon modes are associated with semimetalic and semiconducting bands, respectively.



As clearly seen in this work, our developed analysis method, FDARPES, can offer new routes to investigate the electron-phonon couplings through the non-equilibrium states.
Our work using this method provides direct evidence for the DECP mechanisms responsible for the photo-induced IMTs in Ta$_2$NiSe$_5$. Thus, FDARPES can be used to study many other photo-induced phase transitions by observing how the electron band structure is influenced by the specific phonon-mode.
We also emphasize the versatility of FDARPES.
By using the multiple degrees of freedom in the excitation pulses, we can drive different quasiparticles; for example, circularly-polarized pulses can promote a specific spin population or appropriate mid- and far-infrared wavelength can resonantly excite IR-active phonons.
Furthermore, FDARPES can detect couplings of electrons to any quasiparticles or collective modes as long as their couplings manifest as oscillations of intensities in the TARPES spectra.


\begin{acknowledgments}
We would like to acknowledge Y. Ohta, for valuable discussions and comments.
This work was supported by Grants-in-Aid for Scientific Research (KAKENHI) (Grant No. JP18K13498, JP19H00659, JP19H01818, JP19H00651 JP18K14145, and JP19H02623) from the Japan Society for the Promotion of Science (JSPS), by JSPS KAKENHI on Innovative Areas “Quantum Liquid Crystals” (Grant No. JP19H05826), by the Center of Innovation Program from the Japan Science and Technology Agency, JST, the Research and Education Consortium for Innovation of Advanced Integrated Science by JST, and by MEXT Quantum Leap Flagship Program (MEXT Q-LEAP) (Grant No. JPMXS0118067246, JPMXS0118068681), Japan.
The computation in this work has been done using the facilities of the Supercomputer Center, The Institute for Solid State Physics, The University of Tokyo.
\end{acknowledgments}

\clearpage
\onecolumngrid
\section{Supplemental Material}
\subsection{Physical description and simulation scheme}
Most coherent phonons excited by displacive excitation of coherent phonon (DECP) mechanisms [S1] show a similar phonon frequency to ground state frequency evaluated with Raman scattering measurements [S2-S5]. The frequency-domain ARPES (FDARPES) spectra shown in Figs. 4(a)-4(e) are attributed to the Fourier transform of ARPES modulation owing to the coherent phonon excited by the pump pulse.
To identify the phonon modes, we extracted electronic structure modulation due to phonon modes from theoretical simulations.
We employed an ab-initio theoretical framework based on density-functional theory (DFT) [S6, S7].
We evaluated the phonon modes and take all-symmetric modes, $A_g$ irreducible representation, from the framework.

\subsection{Details of theoretical simulation}
Our theoretical simulation relied on the Perdew–Burke-Ernzerhof (PBE) functional [S8] within density-functional theory (DFT).
Whole calculations were performed by the ABINIT code [S9].
We used the fhi98PP pseudopotential [S10] for abinit [S11] with plane-wave basis set whose energy cut-off was chosen as 50 Hartree, 1361 eV.
To enforce the proper crystal spatial symmetry, C2/c (No. 15), we had to use the Bravais lattice as the simulation cell rather than the primitive cell.
The supercell contains two primitive cells. We employ $24\times6\times6$ Brillouin zone sampling through out whole calculations unless we specifically denote a condition.

\subsection{Atomic position optimization}
First, we performed atomic position optimization based on PBE-DFT with the same cell shape as an experimental value, $a$ = 3.496 \AA, $b$ = 12.829 \AA, $c$ = 15.641 \AA, $\alpha$ = $\gamma$ = $90^\circ
$, $\beta$ = $90.53^\circ$ [S12].
The tolerance of the optimization was $5.0 \times 10^{-5}$ a.u. = 0.0257 eV nm$^{-1}$. The DFT-optimal atomic and experimental positions are shown in Table S1 and Table S2.
The input file for SCF within ABINIT is attached as supplemental material (input\_SCF.in).

\twocolumngrid

\begin{table}[!h]
\caption{\label{S1}Atomic positions of the optimal crystal structure from DFT}
\begin{ruledtabular}
\begin{tabular}{cccc}
 & $x$ & $y$ & $z$ \\
\hline
Ta & -0.004440 &0.220835 & 0.109287 \\
Ni & 0.0 & 0.701355 & 1/4 \\
Se(1) & 0.5046 & 0.077651 & 0.138568 \\
Se(2) & -0.003119 & 0.142849 & 0.950279 \\
Se(3) & 0.0  & 0.32910 & 1/4
\end{tabular}
\end{ruledtabular}
\end{table}

\begin{table}[!h]
\caption{\label{S2}Atomic positions of the experimental crystal structure}
\begin{ruledtabular}
\begin{tabular}{cccc}
 & $x$ & $y$ & $z$ \\
\hline
Ta & -0.007930 & 0.221349 & 0.110442 \\
Ni & 0.0 & 0.701130 & 1/4 \\
Se(1) & 0.5053 & 0.080385 & 0.137979 \\
Se(2) & -0.005130 & 0.145648 & 0.950866 \\
Se(3) & 0.0 & 0.32714 & 1/4
\end{tabular}
\end{ruledtabular}
\end{table}

\onecolumngrid

\subsection{Phonon mode calculation and its assignment}
Second, we obtained phonon modes based on density-functional perturbation theory (DFPT) [S13]. The normal mode $U_{a,\alpha}^q$ for atom a was obtained as an eigenfunction of the dynamical matrix through DFPT as
\begin{equation}
\sum_{b\beta} \mathcal{D}_{a\alpha;b\beta} U_{b, \beta}^{q} = \omega^{2}(q) U_{a, \alpha}^{q},\ 
\alpha,\beta = x,y,z,\ 
\sum_{a\alpha} \left|U^{q}_{a, \alpha} \right|^2 =1
\end{equation}
where $q$ is an index for the eigenmode.
The input file for DFPT within ABINIT is attached as supplemental material (input\_DFPT.in).
The actual atomic displacement was obtained by $\delta R_{a,\alpha}^q = U_{a,\alpha}^q/\sqrt{M_a}$, using mass $M_a$ of an atom $a$.
Because the Bravais lattice is a double size supercell, the phonon modes obtained by DFPT contain $\Gamma$-point phonons and finite wavenumber phonons.
Since the equivalent atomic displacements for the primitive cell of $\Gamma$-point modes is in-phase, the relative motions of the equivalent atoms are the reference to extract the relevant phonon modes.
We rejected modes that showed counter displacements between two-equivalent atoms for the primitive cell.

We identified an irreducible representation for each mode by performing symmetry operations for the atomic displacement and evaluating characters by symmetry operation for the general points of the Bravais lattice,
\begin{equation}
R_1 \left(\begin{array}{c} x_1 \\ x_2 \\x_3 \end{array} \right) = \left(\begin{array}{c} x_1 \\ x_2 \\ x_3 \end{array} \right),\ 
R_2 \left(\begin{array}{c} x_1 \\ x_2 \\x_3 \end{array} \right) = \left(\begin{array}{c} -x_1 \\ x_2 \\ -x_3+1/2 \end{array} \right),\ 
R_3 \left(\begin{array}{c} x_1 \\ x_2 \\x_3 \end{array} \right) = \left(\begin{array}{c} -x_1 \\ -x_2 \\ -x_3 \end{array} \right),\ 
R_4 \left(\begin{array}{c} x_1 \\ x_2 \\x_3 \end{array} \right) = \left(\begin{array}{c} x_1 \\ -x_2 \\ x_3+1/2 \end{array} \right)
\end{equation}
and character table, Table S3.

\begin{table}[!h]
\caption{\label{S3}Characters table of the C2/c space group}
\begin{ruledtabular}
\begin{tabular}{ccccc}
 & $R_1$ & $R_2$ & $R_3$ & $R_4$ \\
\hline
$A_g$ & 1 & 1 & 1 & 1 \\
$A_u$ & 1 & 1 & -1 & -1 \\
$B_g$ & 1 & -1 & 1 & -1 \\
$B_u$ & 1 & -1 & -1 & 1
\end{tabular}
\end{ruledtabular}
\end{table}

Numerically evaluated characters for the displacement were evaluated as
\begin{equation}
\mathcal{C}_{i}^{q} = \sum_a \left[ \left( \delta r_{b,1}^q, \delta r_{b,2}^q, \delta r_{b,3}^q \right) \tilde{R}_i  \left(\begin{array}{c} \delta r_{a,1}^q \\ \delta r_{a,2}^q \\ \delta r_{a,3}^q \end{array} \right) \right] \bigg/ \sum_a \left[ \right( \delta r_{a,1}^{q}\left)^2 + \right( \delta r_{a,2}^{q}\left)^2 + \right( \delta r_{a,3}^{q}\left)^2 \right]
\end{equation}
where $\delta r_{a}^{q}$ is the coordinate for the lattice vectors rather than Cartesian coordinate, atom $b$ is that transformed by $R_i$ from atom $a$, and $\tilde{R}_i$ is a translation-free symmetry operation as
\begin{equation}
\tilde{R}_1 \left(\begin{array}{c} x_1 \\ x_2 \\x_3 \end{array} \right) = \left(\begin{array}{c} x_1 \\ x_2 \\ x_3 \end{array} \right),\ 
\tilde{R}_2 \left(\begin{array}{c} x_1 \\ x_2 \\x_3 \end{array} \right) = \left(\begin{array}{c} -x_1 \\ x_2 \\ -x_3 \end{array} \right),\ 
\tilde{R}_3 \left(\begin{array}{c} x_1 \\ x_2 \\x_3 \end{array} \right) = \left(\begin{array}{c} -x_1 \\ -x_2 \\ -x_3 \end{array} \right),\ 
\tilde{R}_4 \left(\begin{array}{c} x_1 \\ x_2 \\x_3 \end{array} \right) = \left(\begin{array}{c} x_1 \\ -x_2 \\ x_3 \end{array} \right)
\end{equation}
because the displacement itself is a translation-free object. Errors of the numerical character were smaller than $1.0 \times 10^{-5}$ compared with the expected unity.

We focused on the $A_g$ mode because the DECP mechanism informs us that the all-symmetric $A_g$ mode is excited.
The frequencies are shown in Table S4.
Actual modes are attached as supplemental material (PBE\_phonon.csv).

\begin{table}[!h]
\caption{\label{S4}Frequencies of $A_g$ mode, quasicharacters and assigned experimental modes}
\begin{ruledtabular}
\begin{tabular}{ccccccc}
Frequency(THz) & Quasicharacter for & Assigned to & C2 & C4 & C6 &C8 \\
 & orthorhombic phase & experimental mode [S5] & & & & \\ 
\hline
0.99 & $A_g$ & 1.08 THz & 0.998 & 0.998 & 0.998 & 0.998 \\
1.96 & $B_{2g}$ & 2.14 THz & -1.000 & -1.000 & -1.000 & -1.000 \\
2.42 & $B_{2g}$ & 3.71 THz & -0.979 & -0.979 & -0.979 & -0.979 \\
2.88 & $A_g$ & 3.04 THz & -0.709 & -0.709 & -0.709 & -0.709 \\
3.57 & $A_g$ & 4.07 THz & 0.700 & 0.700 & 0.700 & 0.700 \\
4.46 & $B_{2g}$ & 4.46 THz & -0.998 & -0.998 & -0.998 & -0.998 \\
5.16 & $A_g$ & 5.38 THz & 0.906 & 0.906 & 0.906 & 0.906 \\
5.68 & $A_g$ & 5.84 THz & 0.830 & 0.830 & 0.830 & 0.830 \\
6.28 & $A_g$ & 6.52 THz & 0.780 & 0.780 & 0.780 & 0.780 \\
6.76 & $A_g$ & 7.09 THz & 0.845 & 0.845 & 0.845 & 0.845 \\
8.55 & $A_g$ & 8.78 THz & 0.995 & 0.995 & 0.995 & 0.995
\end{tabular}
\end{ruledtabular}
\end{table}

We assigned theoretical modes to an experimental frequency by the approximated symmetry of the oscillation.
In Table S5, quasicharacters for the orthorhombic phase are shown as well.
By increasing the temperature, the Ta$_2$NiSe$_5$ crystal adopted an orthorhombic crystal structure, Cmcm (No. 63), whose operation and character table are given in Eq. (S5) and Table S5, respectively.
We numerically evaluated the characters based on the Cmcm symmetry and used its value to assign the phonon modes to whether $A_g$ or $B_{2g}$ modes.
This analysis was not rigorously well defined and is an approximated treatment.
We called the numerically evaluated character for the higher symmetry space group as the quasicharacter here.

\begin{gather}
C_1 \left(\begin{array}{c} x_1 \\ x_2 \\x_3 \end{array} \right) = \left(\begin{array}{c} x_1 \\ x_2 \\ x_3 \end{array} \right),\ 
C_2 \left(\begin{array}{c} x_1 \\ x_2 \\x_3 \end{array} \right) = \left(\begin{array}{c} -x_1 \\ -x_2 \\ x_3+1/2 \end{array} \right),\ 
C_3 \left(\begin{array}{c} x_1 \\ x_2 \\x_3 \end{array} \right) = \left(\begin{array}{c} -x_1 \\ x_2 \\ -x_3+1/2 \end{array} \right),\ 
C_4 \left(\begin{array}{c} x_1 \\ x_2 \\x_3 \end{array} \right) = \left(\begin{array}{c} x_1 \\ -x_2 \\ -x_3 \end{array} \right), \notag \\
C_5 \left(\begin{array}{c} x_1 \\ x_2 \\x_3 \end{array} \right) = \left(\begin{array}{c} -x_1 \\ -x_2 \\ -x_3 \end{array} \right),\ 
C_6 \left(\begin{array}{c} x_1 \\ x_2 \\x_3 \end{array} \right) = \left(\begin{array}{c} x_1 \\ x_2 \\ -x_3+1/2 \end{array} \right),\ 
C_7 \left(\begin{array}{c} x_1 \\ x_2 \\x_3 \end{array} \right) = \left(\begin{array}{c} x_1 \\ -x_2 \\ x_3+1/2 \end{array} \right),\ 
C_8 \left(\begin{array}{c} x_1 \\ x_2 \\x_3 \end{array} \right) = \left(\begin{array}{c} -x_1 \\ x_2 \\ x_3 \end{array} \right).
\end{gather}

\begin{table}[!h]
\caption{\label{S5}Characters table of the Cmcm space group}
\begin{ruledtabular}
\begin{tabular}{ccccccccc}
 & $C_1$ & $C_2$ & $C_3$ & $C_4$ & $C_5$ & $C_6$ & $C_7$ & $C_8$ \\
\hline
$A_g$ & 1 & 1 & 1 & 1 & 1 & 1 & 1 & 1 \\
$A_u$ & 1 & 1 & 1 & 1 & -1 & -1 & -1 & -1 \\
$B_{1g}$ & 1 & 1 & -1 & -1 & 1 & 1 & -1 & -1 \\
$B_{1u}$ & 1 &  1 & -1 & -1 & -1 & -1 & 1 & 1 \\
$B_{3g}$ & 1 & -1 & -1 & -1 & 1 & -1 & -1 & 1 \\
$B_{3u}$ & 1 & -1 & -1 & 1 & -1 & 1 & 1 & -1 \\
$B_{2g}$ & 1 & -1 & 1 & -1 & 1 & -1 & 1 & -1 \\
$B_{2u}$ & 1 & -1 & 1 & -1 & -1 & 1 & -1 & 1
\end{tabular}
\end{ruledtabular}
\end{table}

To obtain the quasicharacter, we introduced an approximated identification of the atomic position related to Eq. (S5).
We allowed error value $\epsilon$=0.1 Bohr = 0.00592 nm for the identification as
\begin{equation}
\sqrt{\left[\sum_{p \alpha} A_{\alpha, p} \left(C_i r_{a, \alpha} - r_{b, \alpha} \right) \right]^2} < \epsilon
\end{equation}
where $r_{a,\alpha}$ is the reduced coordinate of atom $a$ and $A$ is a matrix to convert the reduced coordinate to a Cartesian coordinate.
The error value was chosen such that a one-to-one correspondence of atoms was realized for the whole operation given by Eq. (S5).
Although the fourth $A_g$ mode, 2.88 THz, had an almost $B_{2g}$-like character, we assigned it as an $A_g$ mode.
In the orthorhombic phase, the number of $B_{2g}$ mode is three.
We assign three modes, 1.96 THz, 2.42 THz, and 4.46 THz, that had the closest quasicharacter value compared to minus unity as the $B_{2g}$ mode and do not 2.88 THz mode.

\subsection{Follow-up}
To check robustness of our conclusion with this specific framework based on PBE-DFT, we performed calculations in different conditions, 1) Perdew-Wang (PW) functional [S14] was used rather than PBE, 2) PBE-DFT framework but with experimental atomic positions, and 3) PW-DFT framework with experimental atomic positions.
We use LDA pseudopotential [S15] for the PW-DFT calculation.
The atomic position optimized with the PW functional is shown in Table S6.
To obtain the identification of approximately equivalent atoms for the orthorhombic phase, $\epsilon$ = 0.2 Bohr was employed in Eq. (S6) for the experimental crystal structure.
The phonon frequencies and the quasicharacters for these calculation conditions are summarized in Table 
S7.

\begin{table}[!h]
\caption{\label{S6}Atomic position of DFT with the PW functional}
\begin{ruledtabular}
\begin{tabular}{cccc}
 & $x$ & $y$ & $z$ \\
\hline
Ta &-0.007407 & 0.223358 & 0.112221 \\
Ni & 0.0 & 0.7049391 & 1/4 \\
Se(1) & 0.5047 & 0.077647 & 0.138656 \\
Se(2) & -0.004972 & 0.149029 & 0.952662 \\
Se(3) & 0.0 & 0.32730 & 1/4 
\end{tabular}
\end{ruledtabular}
\end{table}

\begin{table}[!h]
\caption{\label{S7}Frequencies of $A_g$ mode, quasicharacters and assigned experimental modes}
\begin{ruledtabular}
\begin{tabular}{cccccc}
Frequency(THz) & Quasicharacter & Frequency (THz) & Quasicharacter & Frequency (THz) & Quasicharacter \\
PW DFT & PW DFT & PBE Exp. & PBE Exp. & PW Exp. & PW Exp. \\
\hline
0.97 & $A_g$ & 0.96 & $A_g$ & 1.05 & $A_g$ \\
1.91 & $B_{2g}$ & 2.03 & $B_{2g}$ & 1.78 & $B_{2g}$ \\
2.88 & $A_g$ & 2.83 & $A_g$ & 2.61 & $B_{2g}$ \\
3.14 & $B_{2g}$ & 3.53 & $B_{2g}$ & 2.85 & $A_{g}$ \\
3.72 & $A_{g}$ & 3.59 & $A_g$ & 3.65 & $A_g$ \\
4.57 & $B_{2g}$ & 4.64 & $B_{2g}$ & 4.29 & $B_{2g}$ \\
5.16 & $A_g$ & 5.32 & $A_g$ & 5.01 & $A_g$ \\
5.67 & $A_g$ & 5.89 & $A_g$ & 5.38 & $A_g$ \\
6.21 & $A_g$ & 6.45 & $A_g$ & 6.01 & $A_g$ \\
6.92 & $A_g$ & 7.04 & $A_g$ & 6.56 & $A_g$ \\
8.80 & $A_g$ & 8.86 & $A_g$ & 8.38 & $A_g$
\end{tabular}
\end{ruledtabular}
\end{table}

Here, we comment on a trade-off between the computational cost and convergence accuracy of the results of this simulation.
The most computationally demanding part of this procedure was the DFPT part because many and large linear systems must be solved.
When 384 cores of the system-B (Sekirei), supercomputer at the Institute of Solid State Physics, the University of Tokyo, were used within the flat-MPI parallelization, the DFPT calculation took 65.13 hours.
Roughly speaking, the DFPT cost with the GGA-functional obeys a linear scaling for the number of $k$-points.
An estimation for $36 \times 9 \times 9$ Brillouin zone sampling is 21 hours, with 3,600 cores.
This calculation is possible in principle but hardly achievable with the supercomputer system.

To evaluate the convergence accuracy with respect to the density of Brillouin zone sampling, we estimated the error through the residual error of the force and equilibrium atomic position for denser Brillouin zone sampling, $N_K =36 \times 9 \times 9$.
The residual force, evaluated with the optimal atomic position of sparser Brillouin zone sampling, was maximally $1.3 \times 10^{-4}$ = 0.068 eV nm$^{-1}$, which was fairly close to our criteria to identify the atomic equilibrium position.
The optimized atomic positions with denser sampling are shown in Table S8.
By converting these to a Cartesian coordinate, maximal error is $6.8 \times 10^{-6}$ (a.u.) = 0.36 fm.
These errors were small compared with the following case for the energy cut off.

We also checked the energy convergence error by using energy cut off 60 a.u. = 1633 eV with the same Brillouin zone sampling, $N_K = 24 \times 6 \times 6$.
The residual force, evaluated with ecut = 50 a.u., is maximally $2.2 \times 10^{-4}$ a.u. = 0.11 eV nm$^{-1}$. The optimized atomic positions are shown in Table S9.
The maximal error of the atomic position in the Cartesian coordinate was $1.3 \times 10^{-5}$ a.u. = 0.69 fm.
Performing DFPT calculation for the energy cut off 60 a.u., we obtained the phonon frequencies and quasicharacter shown in Table S10.
The error for the frequency of the phonon modes was maximally 0.02 THz.
Thus, the error in the Brillouin zone sampling is expected to be smaller than this error.

\twocolumngrid

\begin{table}[!h]
\caption{\label{S8}Atomic positions of optimal crystal structure from PBE-DFT with $36 \times 9 \times 9$ Brillouin zone sampling}
\begin{ruledtabular}
\begin{tabular}{cccc}
 & $x$ & $y$ & $z$ \\
\hline
Ta & -0.004442 & 0.220801 & 0.109289 \\
Ni & 0.0 & 0.7013654 & 1/4 \\
Se(1) & 0.5046 & 0.077653 & 0.138552 \\
Se(2) & -0.003147 & 0.142847 & 0.950286 \\
Se(3) & 0.0 & 0.32910 & 1/4
\end{tabular}
\end{ruledtabular}
\end{table}

\begin{table}[!h]
\caption{\label{S9}Atomic positions of optimal crystal structure from PBE-DFT with energy cut off 60 a.u.}
\begin{ruledtabular}
\begin{tabular}{cccc}
 & $x$ & $y$ & $z$ \\
\hline
Ta & -0.004440 & 0.220819 & 0.109280 \\
Ni & 0.0 & 0.7014205 & 1/4 \\
Se(1) & 0.5046 & 0.077647 & 0.138557 \\
Se(2) & -0.003124 & 0.142859 & 0.950281 \\
Se(3) & 0.0 & 0.32910 & 1/4
\end{tabular}
\end{ruledtabular}
\end{table}

\newpage
\onecolumngrid

\begin{table}[!h]
\caption{\label{S10}Frequencies of optimal crystal structure from PBE-DFT energy cut off 60 a.u.}
\begin{ruledtabular}
\begin{tabular}{cccccc}
Frequency(THz) & Quasicharacter for &  C2 & C4 & C6 &C8 \\
 & orthorhombic phase & & & \\
\hline
0.97 & $A_g$ & 0.998 & 0.998 & 0.998 & 0.998 \\
1.96 & $B_{2g}$ & -1.000 & -1.000 & -1.000 & -1.000 \\
2.42 & $B_{2g}$ & -0.979 & -0.979 & -0.979 & -0.979 \\
2.87 & $A_g$ & -0.713 & -0.713 & -0.713 & -0.713 \\
3.56 & $A_g$ & 0.684 & 0.684 & 0.684 & 0.684 \\
4.46 & $B_{2g}$ & -0.998 & -0.998 & -0.998 & -0.998 \\
5.16 & $A_g$ & 0.900 & 0.900 & 0.900 & 0.900 \\
5.67 & $A_g$ & 0.830 & 0.830 & 0.830 & 0.830 \\
6.28 & $A_g$ & 0.780 & 0.780 & 0.780 & 0.780 \\
6.76 & $A_g$ & 0.844 & 0.844 & 0.844 & 0.844 \\
8.53 & $A_g$ & 0.995 & 0.995 & 0.995 & 0.995 
\end{tabular}
\end{ruledtabular}
\end{table}

\newpage

\subsection{Correlation between the orthorhombic and monoclinic phases}
By reducing the symmetry from the orthorhombic to monoclinic phase, the number of irreducible representations is also reduced.
As a result, Raman-active phonon modes of both $A_{g}$ and $B_{2g}$ characters in the orthorhombic phase turn to be $A_{g}$ character in the monoclinic phase.
Tables S11 and S12 show Wyckoff positions and Raman-active phonon modes for each atom in the orthorhombic (Cmcm) and monoclinic (C2/c) phases, respectively.
Table S13 shows correlation for each irreducible representation between the orthorhombic (Cmcm) and monoclinic (C2/c) phase, and Table S14 shows the number of total Raman-active phonon modes and its correlation relations.

\twocolumngrid

\begin{table}[!h]
\caption{\label{S11}Wyckoff positions and Raman-active phonon modes for each atom in the orhorhombic (Cmcm) phase}
\begin{ruledtabular}
\begin{tabular}{ccc}
 & Wyckoff position & Raman-active phonon modes \\
\hline
Ta & 8f & $2A_{g}+B_{1g}+B_{2g}+2B_{3g}$ \\
Ni & 4c & $A_{g}+B_{1g}+B_{3g}$ \\
Se(1) & 8f & $2A_{g}+B_{1g}+B_{2g}+2B_{3g}$ \\
Se(2) & 8f & $2A_{g}+B_{1g}+B_{2g}+2B_{3g}$ \\
Se(3) & 4c & $A_{g}+B_{1g}+B_{3g}$
\end{tabular}
\end{ruledtabular}
\end{table}

\begin{table}[!h]
\caption{\label{S12}Wyckoff positions and Raman-active phonon modes for each atom in the monoclinic (C2/c) phase}
\begin{ruledtabular}
\begin{tabular}{ccc}
 & Wyckoff position & Raman-active phonon modes \\
\hline
Ta & 8f & $3A_{g}+3B_{g}$ \\
Ni & 4e & $A_{g}+2B_{g}$ \\
Se(1) & 8f & $3A_{g}+3B_{g}$ \\
Se(2) & 8f & $3A_{g}+3B_{g}$ \\
Se(3) & 4e & $A_{g}+2B_{g}$
\end{tabular}
\end{ruledtabular}
\end{table}

\begin{table}[!h]
\caption{\label{S13}Correlation table between the orhorhombic (Cmcm) and monoclinic (C2/c) phase}
\begin{ruledtabular}
\begin{tabular}{cc}
Cmcm & C2/c \\
\hline
$A_{g}$ & $A_{g}$ \\
$A_{u}$ & $A_{u}$ \\
$B_{1g}$ & $B_{g}$ \\
$B_{1u}$ & $B_{u}$ \\
$B_{3g}$ & $B_{g}$ \\
$B_{3u}$ & $B_{u}$ \\
$B_{2g}$ & $A_{g}$ \\
$B_{2u}$ & $A_{u}$
\end{tabular}
\end{ruledtabular}
\end{table}

\begin{table}[!h]
\caption{\label{S14}The number of total Raman-active phonon mondes and its correlation relations}
\begin{ruledtabular}
\begin{tabular}{cc}
Cmcm & C2/c \\
\hline
$8A_{g} + 3B_{2g}$ & $11A_{g}$ \\
$5B_{1g} + 8B_{3g}$ & $13B_{g}$
\end{tabular}
\end{ruledtabular}
\end{table}

\newpage

\onecolumngrid
\subsection{Comparison between band dispersions and FDARPES spectra}
We compare band dispersions before and after photoexcitation with FDARPES spectra for each frequency as shown in Fig. S1.

\renewcommand\thefigure{S\arabic{figure}} 
\setcounter{figure}{0}    
\begin{figure}[!h]
\includegraphics{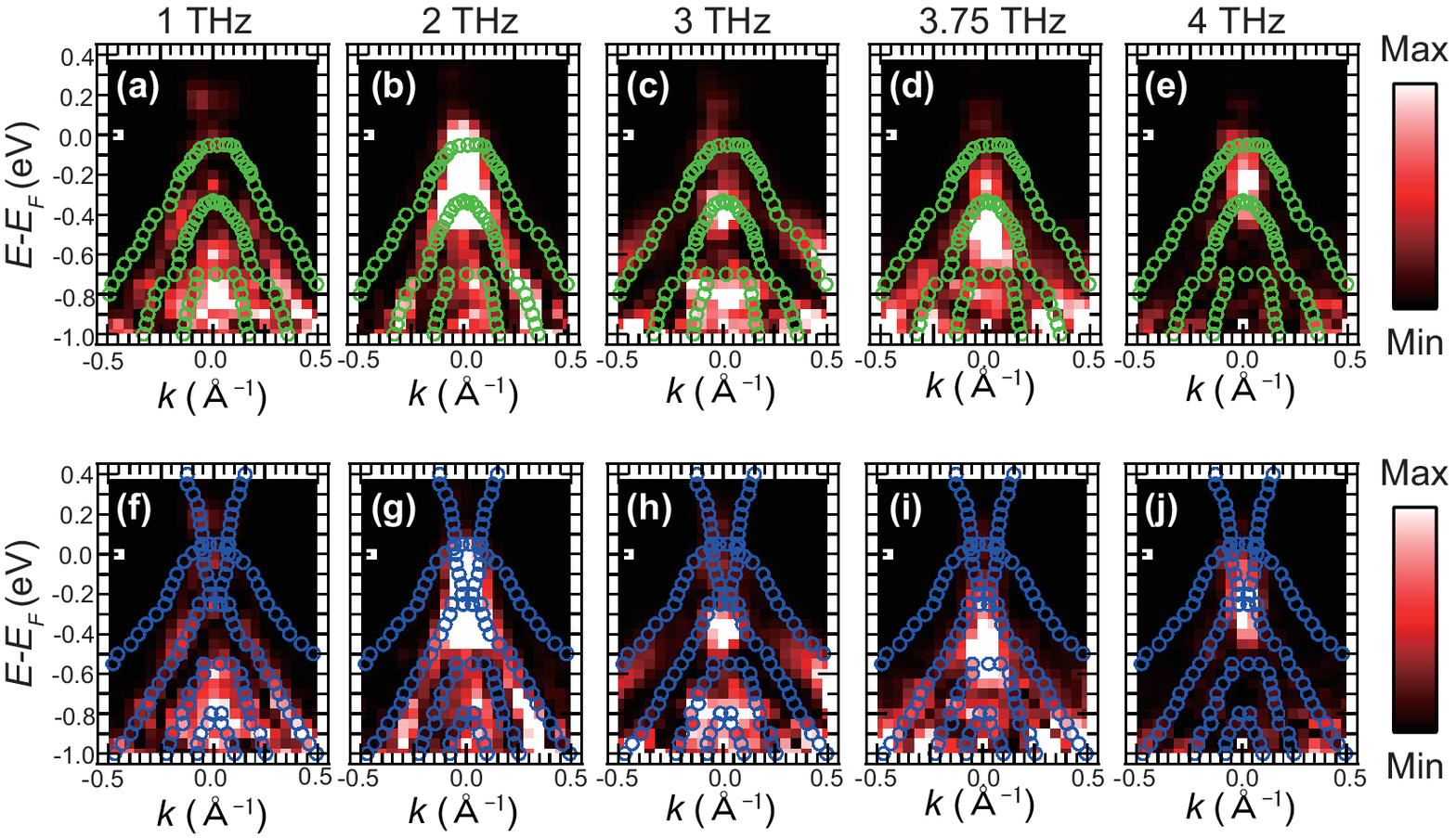}
\caption{
Comparison between FDARPES spectra and band dispersions before (Figs. S1(a)-S1(e)) and after photoexcitation (Figs. S1(f)-S1(j)).
Band dispersions are shown as green (Figs. S1(a)-S1(e)) and blue circles (Figs. S1(f)-S1(j)).
}
\label{figS1}
\end{figure}

\newpage

\section{References}
\begin{enumerate}
\renewcommand{\labelenumi}{[S\arabic{enumi}]}
\item Zeiger, H. J., Vidal, J., Cheng, T. K., Ippen, E. P., Dresselhaus, G., and Dresselhaus, M. S. Phys. Rev. B \textbf{45}, 768 (1992).
\item Misochko, O. V. and Lebedev, M. V. J. Exp. Theor. Phys. \textbf{126}, 64 (2018).
\item Garrett, G. A., Albrecht, T. F., Whitaker, J. F., and Merlin. R. Phys. Rev. Lett. \textbf{77}, 3661 (1996).
\item Ishioka, K., Kitajima, M., and Misochko, O. V. J. Appl. Phys. \textbf{103}, 123505 (2008).
\item Werdehausen, D., Takayama, T., Höppner, M., Albrecht, G., Rost, A. W., Lu, Y. Manske, D., Takagi, H. and Kaiser, S. Sci. Adv. \textbf{4}, eaap8652 (2018).
\item Hohenberg, P. and Kohn., W. Phys. Rev. \textbf{136}, B864 (1964).
\item Kohn, W. and Sham, L. J. Phys. Rev. \textbf{140}, A1133 (1965).
\item Perdew, J. P., Burke, K., and Ernzerhof, M. Phys. Rev. Lett. \textbf{77}, 3865 (1996).
\item Gonze, X. et al., Comput. Phys. Commun. \textbf{205}, 106-131 (2016).
\item Fuchs, M. and Scheffler, M. Comput. Phys. Commun. \textbf{119}, 67 (1999).
\item \url{https://www.abinit.org/sites/default/files/PrevAtomicData/psp-links/gga_fhi.html}
\item Sunshine, S. A. and Ibers, J. A. Inorg. Chem. \textbf{24}, 3611 (1985).
\item de Gironcoli, B. S., Dal Corso, A. and Giannozi, P. Rev. Mod. Phys. \textbf{73}, 515 (2001).
\item Perdew, J. P., and Wang, Y. Phys. Rev. B \textbf{45}, 13244 (1992).
\item \url{https://www.abinit.org/sites/default/files/PrevAtomicData/psp-links/lda_fhi.html}
\end{enumerate}

\end{document}